\documentstyle[art12]{article}
\date{}
\begin{document}
\title{{\bf
%
Equations and Integrals of Motion in
Discrete Integrable  $A_{k-1}$ Algebra Models
}}

\author{{\large A. P. Protogenov$^{\dagger}$ and V. A. Verbus$^{\ddagger}$}}
\maketitle
\begin{center}
{\footnotesize {\em
$^{\dagger}$Institute of Applied Physics, Russian Academy of Sciences,\\
46 Ul'yanov Street, Nizhny Novgorod 603600, Russia \\
$^{\ddagger}$Institute for Physics of Microstructures, Russian Academy of Sciences,\\
46 Ul'yanov Street, GSP-105, Nizhny Novgorod 603600, Russia}}
\end{center}

\begin{abstract}
We study integrals of motion for Hirota bilinear difference equation
that is satisfied by the eigenvalues of the transfer-matrix.
The combinations of the eigenvalues of the transfer-matrix are found,
which are integrals of motion for integrable discrete models
for the $A_{k-1}$ algebra with zero and quasiperiodic
boundary conditions. Discrete analogues of the equations of motion
for the Bullough-Dodd model and non-Abelian generalization
of Liouville model are obtained.\\\\

\noindent
{\em PACS}: 03.20.+i, 11.30.-j, 11.10.Lm \\
{\em Keywords}: Discrete integrable models; Integrals of motion;
Hirota equation; Transfer-matrix
\end{abstract}
\section*{1. Introduction}

Thermodynamical Bethe-anzatz (TBA) \cite{Zam} is one
of the main methods of studying
statistical properties of integrable low-dimensional systems.
Its formulation yields the functional equations \cite{Kl,Kun,Rav}
which have the form of multidimensional recurrent relations.
The same equations (identical in form but different in the
analytical properties of solutions) are known in the theory
of nonlinear waves as the Hirota equations.
Discrete Hirota equations with specified boundary conditions may yield all
the known exactly integrated (1+1)D equations at the continuous limit.
The problem of integrability of discrete equations was
discussed in recent papers \cite{Wieg1,Zabr,Krich,Lip,Suris,Nih,Cap}.
In Refs. \cite{Wieg1,Zabr,Krich,Lip,Wieg} a significant progress was
achieved in applying Bethe-ansatz (BA) for integrable discrete (1+1)D
systems with the internal degrees of freedom (see also \cite{Kor}).
It was found out thereby that the discrete time variable and
the variable corresponding to the internal degrees of freedom, as well
as discrete rapidities enter
the Hirota equation as the arguments of the searched function on equal
footing.
The limit of continuous time in this case corresponds to
large spin values \cite{Fadd}.

The behavior of the considered system is
universal due to several reasons. The formulation of the model on the
lattice leads to the solutions \cite{Krich}, which are expressed
via elliptic functions. Complexification of the parameters includes
the possibility to consider the quantum-group situation with the deformation
parameter being a root of unity \cite{Resh,Bern}.
Different values for the shifts of rapidities along the
imaginary axis in the searched functions of the discrete
equations lead to different functional equations appearing
in the TBA approach. In the particular case, when
the rapidities are much greater than these shifts,
we deal with recurrent relations \cite{Prot1,Prot2,Fu}
existing in the theory that takes into account the Haldane's
generalized principle of exclusion statistics \cite{Hal}.

The eigenvalues $T_{s}^{a}(u)$ of the transfer matrix,
satisfying Eq.(\ref {E5}) with certain boundary conditions
(see below),
are characterized by analytical properties of solutions
specific for the problem under consideration.
Thus, for example, the functional recurrent
relations of a thermodynamical Bethe-anzatz \cite{Rav},
being outwardly similar in form with equations
for $A_{1}-$case with zero boundary condition,
are satisfied by functions with different analytical
properties. The distinction in analytical properties
consists in the fact that for recurrent TBA equations
the shift of rapidities occurs along the imaginary axis.

In the present paper we consider the case of real shifts
along the axis of rapidities. The main goal of the paper is search
for integrals of motion following from the analysis of the
Hirota classical equation for zero and quasiperiodic boundary
conditions in the case of $A_{k-1}$ algebra.
The integrals of motion are useful in seeking solutions
of BA equations.
As the example of the equations that require integrals
of motion to be analyzed we consider equations in the discrete
and continuous limits for the $k = 3$ case.
The general approach for studying the integrals
of motion for discrete sine-Gordon model was used in
Ref. \cite{Feigin}. The representation of the zero
curvature \cite{Fadd2,Zah} for the classical discrete sine-Gordon
model via the R-matrix
was studied in Ref. \cite{Za}.

The present paper is organized as follows. In the second
section for consistency we provide the main expressions from Refs.
\cite{Wieg,Saito}. Based on these equations we write out
the discrete equations of motion for the non-Abelian case $k =3$
under zero and quasiperiodic boundary conditions.
Here we give their form in the continuous limit.
The third section is devoted to the analysis of integrals of motion
in the case of quasiperiodic and zero boundary conditions. In the
discussion we analyze the form of solutions for the
$A_{1}$ algebra with the quasiperiodic boundary conditions.

\section*{2. Hirota Equation}

The Hirota bilinear difference equation may be written down in
different forms. One of the presentation of this equation for the
function $\tau(n,l,m)$ depending on the discrete
variables $n, l$ and $m$ is
$$
\alpha \tau(n,l+1,m) \tau(n,l,m+1) + \beta \tau(n,l,m) \tau(n,l+1,m+1) +
$$
\begin{equation}
\gamma \tau(n+1,l+1,m)\tau(n-1,l,m+1) = 0 \, .
\label{E1}
\end{equation}
The arbitrary constants $\alpha , \beta , \gamma $ in this equation
are restricted by $\alpha + \beta + \gamma  = 0 $.
Further specification of
Eq.(\ref {E1}) is performed with the aid of boundary conditions
that set up the correspondence with the searched model in the
continuous limit.
Preserving integrability, various methods of the realization
of the continuous limit permit to obtain the whole spectrum of
soliton equations in the continuous case.
The transform of the function $\tau(n,l,m)$,
\begin{equation}
\tau(n,l,m) = \frac{(-\alpha / \gamma )^{n^{2}/2}}
{(1+ \gamma /\alpha)^{lm}}\tau_{n}(l,m)
\label{E2}
\end{equation}
eliminates the arbitrary constants
$\alpha , \beta , \gamma $ in Eq.(\ref {E1}):
\begin{equation}
\tau_{n}(l+1,m) \tau_{n}(l,m+1) - \tau_{n}(l,m) \tau_{n}(l+1,m+1)
+ \tau_{n+1}(l+1,m) \tau_{n-1}(l,m-1) = 0 \, .
\label{E3}
\end{equation}

Substitution of the variables
$$
a = n \, , \, \,    s = l + m \, , \,  \,   u = l - m - n \, ,
$$
\begin{equation}
\tau_{n}(l,m) = T_{l+m}^{n}(l-m-n) = T_{s}^{a}(u)
\label{E4}
\end{equation}
leads to the following bilinear functional relations
(expressing the fusion rules) for the eigenvalues
$T_{s}^{a}(u)$ of the transfer matrix in integrable
lattice models for the $A_{k-1}$ algebra:
\begin{equation}
T_{s}^{a}(u+1)T_{s}^{a}(u-1) - T_{s+1}^{a}(u)T_{s-1}^{a}(u) =
T_{s}^{a+1}(u)T_{s}^{a-1}(u) \, .
\label{E5}
\end{equation}
The discrete index $a = 0, 1, \ldots k$ for the $A_{k-1}$
algebra and the discrete time $s$ in this equation are,
respectively, the length and the height of the Young
rectangular diagram, to which the eigenvalue $T_{s}^{a}(u)$
of the transfer matrix  corresponds;
$u$ in Eq.(\ref {E5}) is the discrete spectral parameter.
Equation (\ref {E5}) preserves its form if
the function $T_{s}^{a}(u)$ is multiplied by the product
of four arbitrary functions $\chi _{i}$ depending on
one variable with the following arguments:
\begin{equation}
T_{s}^{a}(u) \to \chi _{1}(a+u+s) \chi _{2}(a-u+s)
\chi _{3}(a+u-s) \chi _{4}(a-u-s)
T_{s}^{a}(u) \, .
\label{E6}
\end{equation}
For the function
\begin{equation}
Y_{s}^{a}(u)  = \frac{T_{s+1}^{a}(u)T_{s-1}^{a}(u)}
{T_{s}^{a+1}(u)T_{s}^{a-1}(u)} \, ,
\label{E7}
\end{equation}
which is invariant with respect to the transform (\ref {E6}), Hirota
equation (\ref {E5}) has the form
\begin{equation}
Y_{s}^{a}(u+1)Y_{s}^{a}(u-1)  =
\frac{\left(1 + Y_{s+1}^{a}(u)\right)\left(1 + Y_{s-1}^{a}(u)\right)}
{\left(1 + (Y_{s}^{a+1}(u))^{-1}\right)\left(1 +
(Y_{s}^{a-1}(u))^{-1}\right)} \, .
\label{E8}
\end{equation}
The zero boundary conditions for the function $T_{s}^{a}(u)$
are as follows \cite{Wieg}
\begin{equation}
T_{s}^{a}(u) = 0 \, \, for \, \, a<0 \, \, and \, \,  a>k \, ,
\label{E9}
\end{equation}
$$
T_{s}^{k}(u) = \phi (u-s-k)\, , \, \, \,  T_{s}^{0}(u) = \phi (u+s) \, ,
$$
where $\phi (u) = \prod\limits_{i=1}^N \sigma(\eta (u-y_{i}))$,
$\sigma$ is the Weierstrass function, $y_{i}$ being its roots;
$N$ is the size of the system and $\eta$ is a parameter.
In the case of $A_{1}-$ algebra, these boundary conditions
lead to the discrete Liouville equation \cite{Hirot}.
In terms of the gauge-invariant variables it has the form
\begin{equation}
Y_{s}^{a}(u-1)Y_{s}^{a}(u+1)  =
\left(1 + Y_{s+1}^{a}(u)\right)\left(1 + Y_{s-1}^{a}(u)\right)
\label{E10}
\end{equation}
with the boundary condition $Y_{0}(u)=0$. In the continuous
limit after replacing $Y_{s}^{a}(u) = \delta^{-2}\exp (-\phi (x,t)),
n =x/\delta, s =t/\delta $ with $ \delta \to 0$
Eq.(\ref {E10}) transforms to Liouville equation
\begin{equation}
\frac{\partial ^{2}\phi }{\partial t^{2}} -
\frac{\partial ^{2}\phi }{\partial x^{2}} =
2e\,^{\phi } \, .
\label{E11}
\end{equation}
The quasiperiodic boundary conditions for the function
\begin{equation}
T_{s}^{a}(u) = e^{\alpha}
\lambda^{-2a}T_{s}^{a+2}(u+2)
\label{E12}
\end{equation}
in the $A_{1}$-algebra case lead to a discrete sine-Gordon
equation\cite{Wieg} for the gauge-invariant function
$X_{s}^{a}(u) = -\lambda (1+Y_{s}^{a}(u))$
\cite{Hirot2,FaVolk,Volk}:
\begin{equation}
X_{s+1}^{a}(u)X_{s-1}^{a}(u) =
\frac{\left(\lambda +X_{s}^{a}(u+1)\right)
\left(\lambda + X_{s}^{a}(u-1)\right)}
{\left(1+\lambda X_{s}^{a}(u+1)\right)
\left(1+ \lambda X_{s}^{a}(u-1)\right)} \, .
\label{E13}
\end{equation}

Due to the boundary condition (\ref {E12}), the functions
$X_{s+1}^{a}(u)$ with different values
of variable "$a$"  are related as
\begin{equation}
X_{s}^{a}(u)X_{s}^{a+1}(u+1) = 1 \, .
\label{E14}
\end{equation}

We consider discrete equations of motion for the $A_{2}$
algebra for the zero and quasiperiodic boundary conditions,
and their continuous limit. Applying the general approach
of the paper \cite{Wieg} in the case of zero boundary conditions
for gauge-invariant functions $Y_{s}^{a}(u) $ with $a = 1, 2$
we obtain the following system of discrete equations:
\begin{eqnarray}
(1+Y_{s}^{2}(u))Y_{s}^{1}(u+1)Y_{s}^{1}(u+1) =
(1+Y_{s+1}^{1}(u))(1+Y_{s-1}^{1}(u))Y_{s}^{2}(u) \, ,
\label{E15}
\\
(1+Y_{s}^{1}(u))Y_{s}^{2}(u+1)Y_{s}^{2}(u+1) =
(1+Y_{s+1}^{2}(u))(1+Y_{s-1}^{2}(u))Y_{s}^{1}(u) \, .
\label{E16}
\end{eqnarray}
The Eqs.(15),(16) are the first result of this paper.

Similarly to the consideration of the transition from
Eq.(\ref {E10}) to Eq.(\ref {E11}) we now introduce
the functions  $\phi _{l}(x,t)$ with $l = 1, 2 $ in such a way that
$Y_{s}^{1}(u) =\delta ^{-2}\exp(-\phi _{l}(x,t))$.
Let $u=x/\delta, s=t/\delta $.
Then the continuous limit $\delta \to 0$  yields
$SU(3)$ Toda equations \cite{Pi}:
\begin{eqnarray}
\frac{\partial ^{2}\phi _{1}}{\partial t^{2}} -
\frac{\partial ^{2}\phi _{1}}{\partial x^{2}} =
2e\,^{\phi _{1}} - e\,^{\phi _{2}} \, ,
\label{E17}
\\
\frac{\partial ^{2}\phi _{2}}{\partial t^{2}} -
\frac{\partial ^{2}\phi _{2}}{\partial x^{2}} =
2e\,^{\phi _{2}} - e\,^{\phi _{1}} \, .
\label{E18}
\end{eqnarray}

In the case of non-Abelian discrete sine-Gordon model the
quasiperiodic boundary condition (\ref {E12}) should be generalized
for the $A_{k-1}$ algebra in the following way:
\begin{equation}
T_{s}^{a}(u) = e\,^{\alpha}\,
\lambda ^{-ka}T_{s}^{a+k}(u+k) \, .
\label{E19}
\end{equation}
Here the functions $X_{s}^{a}(u)$ satisfy
the relation
\begin{equation}
X_{s}^{a}(u)X_{s}^{a+1}(u+1) \ldots X_{s}^{a+k-1}(u+k-1) = (-1)^{k} \, ,
\label{E20}
\end{equation}
which generalizes Eq.(\ref {E14}).

Consider the case $k = 3$. The Hirota equation for gauge-invariant
functions $X_{s}^{1}(u)$ for the $A_{2}-$algebra with boundary
conditions (\ref {E19}) can be written as
\begin{eqnarray}
(\lambda + X_{s}^{2}(u))X_{s+1}^{1}(u)X_{s-1}^{1}(u) =
\frac{(\lambda +X_{s}^{1}(u-1))(\lambda + X_{s}^{1}(u+1))}
{(1-\lambda X_{s}^{1}(u+1)X_{s}^{2}(u+2))}X_{s}^{2}(u) \, ,
\label{E21}
\\
(\lambda + X_{s}^{1}(u))X_{s+1}^{2}(u)X_{s-1}^{2}(u) =
\frac{(\lambda +X_{s}^{2}(u-1))(\lambda + X_{s}^{2}(u+1))}
{(1-\lambda X_{s}^{1}(u-2)X_{s}^{2}(u-1))}X_{s}^{1}(u) \, .
\label{E22}
\end{eqnarray}
This equations of motion is the second result of our paper.

The continuous limit in these equations can be achieved
as follows. Assume that $\lambda \to 0$ and $s \to \infty $
in such a way that $s = t/\sqrt{\lambda}$  for a fixed $t$.
Similarly, assume that $\lambda \to 0 $ and $u \to \infty $ with
$u = x/\sqrt{\lambda}$  for a fixed $x$.
Being parametrized with the aid of the function
$ X^{l}(x,t) = -\exp(-\phi_{l}(x,t) )$
Eqs. (\ref {E21}) and (\ref {E22}) transform in the continuous
limit to the following equations of motion
\begin{eqnarray}
\frac{\partial ^{2}\phi _{1}}{\partial t^{2}} -
\frac{\partial ^{2}\phi _{1}}{\partial x^{2}} =
2e\,^{\phi _{1}} - e\,^{\phi _{2}} - e\,^{-(\phi _{1}+\phi _{2})} \, ,
\label{E23}
\\
\frac{\partial ^{2}\phi _{2}}{\partial t^{2}} -
\frac{\partial ^{2}\phi _{2}}{\partial x^{2}} =
2e\,^{\phi _{2}} - e\,^{\phi _{1}} - e\,^{-(\phi _{1}+\phi _{2})} \, ,
\label{E24}
\end{eqnarray}
which correspond to $SU(3)-$case of affine Toda equation.

Let us assume that in Eq.(21) $X_{s}^{1}(u) = X_{s}^{2}(u)=X_{s}(u)$.
Then we have
\begin{equation}
\frac{X_{s+1}(u) X_{s-1}(u)}{X_{s}(u)}  =
\frac{(\lambda +X_{s}(u-1))(\lambda + X_{s}(u+1))}
{(1 - \lambda X_{s}(u-2)X_{s}(u-1))(\lambda +X_{s}(u))} \, ,
\label{Q26}
\end{equation}
or from Eq.(22) with the same assumption
\begin{equation}
\frac{X_{s+1}(u) X_{s-1}(u)}{X_{s}(u)}  =
\frac{(\lambda +X_{s}(u-1))(\lambda + X_{s}(u+1))}
{(1 - \lambda X_{s}(u+2)X_{s}(u+1)) (\lambda +X_{s}(u))} \, .
\label{Q27}
\end{equation}
In this case we have two versions of discrete equation.
These equations are symmetrical to each other with respect
to changing the variable $u \to -u$.
One can symmetrize these equations adding together Eqs.(26),
(27)
$$
\frac{X_{s}(u)}{X_{s+1}(u) X_{s-1}(u)}  =
$$
\begin{equation}
\frac
{(\lambda +X_{s}(u))
(1 - (\lambda /2)X_{s}(u-2)X_{s}(u-1) - (\lambda /2)X_{s}(u+2)X_{s}(u+1))
}
{(\lambda +X_{s}(u-1))(\lambda + X_{s}(u+1))}
\, .
\label{Q28}
\end{equation}

The continuous limit of the Eq.(27) yields
\begin{equation}
\frac{\partial ^{2}\phi }{\partial t^{2}} -
\frac{\partial ^{2}\phi }{\partial x^{2}} =
e\,^{\phi } - e\,^{-2\phi } \, ,
\label{Q25}
\end{equation}
which is the Bullough - Dodd equation.

Another discrete variant of the Bullough-Dodd equation
was found from geometrical point of view in Refs. \cite{Bob1}.
Generalizing the viewpoint, we may
conclude also that Eqs. (\ref {E21}) and (\ref {E22})
are the discrete analog of the equations of motion in
$SU(3)$-case of affine Toda model.
To study solutions of the discrete equations of motion
(\ref {E15}) and (\ref {E16}) and (\ref {E21}) (\ref {E22})
with zero (\ref {E9}) and quasiperiodic (\ref {E19}) boundary
conditions, respectively, one has to find the integrals
of motion. We consider this problem in the next section.

\section*{2. Integrals of motion}
\par
Integrals of motion are known to play the key role in the study of
types of motion.
Under integrals of motion we assume such combination of the functions
$T_{s}^{a}(u)$ which doesn't depent on argument $s$ or $u$.
In Ref. \cite{Wieg} integrals of motion
for discrete dynamics were obtained for the $A_{1}$
algebra and zero boundary conditions. A general
determinant representation of integrals of motion was also written out
for the $A_{k}$ algebra with zero boundary conditions. In this section
we shall show (i) how to get integrals of motion in the $A_{2}$ case
from the known integrals for the $A_{1}$ algebra.
In fact, we use this example to show how integrals of motion are
transformed in the B\"acklund flow. Besides, (ii) we write down
some integrals of motion for quasiperiodic boundary conditions.

We shall use the equations of the linear problem, which ensue from
the gauge and dual transformations on the lattice \cite{Saito}
of the values for the variables in question. These equations
are equivalent to those resulting from the zero curvature
condition \cite{Fadd2,Zah}. Using the notations of Ref. \cite{Saito},
the equations of the linear problem are written as follows
\begin{eqnarray}
g_{n}(l,m)\tau_{n}(l+1,m) - g_{n}(l+1,m)\tau_{n}(l,m) =
c^{-1}g_{n-1}(l,m)\tau_{n+1}(l+1,m),
\label{E25}
\\
g_{n-1}(l,m)\tau_{n}(l,m+1) - g_{n-1}(l,m+1)\tau_{n}(l,m) =
cg_{n}(l,m)\tau_{n-1}(l,m+1).
\label{E26}
\end{eqnarray}
Here $c$ is the arbitrary constant. When comparing
Eqs.(\ref {E25}), (\ref {E26}) with the equations of the
linear problem of the paper \cite{Saito},
the transformation (\ref {E2}) and the relation
$c_{+}c_{-}=-\gamma /\alpha $,  $c_{-} = c$
should be taken into account in (\ref {E25}), (\ref {E26}).
This relation is the necessary condition for the function $g_{n}(l,m)$
to satisfy also Hirota equation. The inverse equations of the
linear problem have the form
\begin{eqnarray}
g_{n}(l,m)\tau_{n}(l-1,m) - g_{n}(l-1,m)\tau_{n}(l,m) =
-c^{-1}g_{n-1}(l-1,m)\tau_{n+1}(l,m),
\label{E27}
\\
g_{n}(l,m)\tau_{n+1}(l,m-1) - g_{n}(l,m-1)\tau_{n+1}(l,m) =
-cg_{n+1}(l,m-1)\tau_{n}(l,m).
\label{E28}
\end{eqnarray}

In the terms of the functions $T_{s}^{a}(u)$ and $G_{s}^{a}(u)$,
using the notations from (\ref {E4}), when
$\tau _{n}(l,m) = T_{l+m}^{n}(l-m-n) = T_{s}^{a}(u)$
and $g_{n}(l,m) = G_{l+m}^{n}(l-m-n) = G_{s}^{a}(u)$,
one can write Eqs (\ref {E25} - \ref {E28}) in the following way.
For the direct equations of the linear problem we have
\begin{eqnarray}
G_{s}^{a}(u) T_{s+1}^{a}(u+1) - G_{s+1}^{a}(u+1) T_{s}^{a}(u) =
c^{-1}G_{s}^{a-1}(u+1) T_{s+1}^{a+1}(u) \, ,
\label{E29}
\\
G_{s}^{a-1}(u+1) T_{s+1}^{a}(u-1) - G_{s+1}^{a-1}(u) T_{s}^{a}(u) =
cG_{s}^{a}(u) T_{s+1}^{a-1}(u) \, .
\label{E30}
\end{eqnarray}
Similarly, the inverse equations are written as
\begin{eqnarray}
G_{s}^{a}(u) T_{s-1}^{a}(u-1) - G_{s-1}^{a}(u-1) T_{s}^{a}(u) =
-c^{-1}G_{s-1}^{a-1}(u) T_{s}^{a+1}(u-1) \, ,
\label{E31}
\\
G_{s}^{a}(u) T_{s-1}^{a+1}(u) - G_{s-1}^{a}(u+1) T_{s}^{a+1}(u-1) =
-cG_{s-1}^{a+1}(u) T_{s}^{a}(u) \, .
\label{E32}
\end{eqnarray}
To make the comparison of the approaches and notations of
various papers \cite{Wieg} easier, we write out the equations
of the linear problems having expressed the functions
included in them via the eigenvalues of the transfer matrix,
used in Ref.\cite{Wieg}. We note that
the complete equivalence is achieved at $c=-1$
and the substitution $s \to -s$.

Let us focus our attention on the symmetry of pairs of equations
(\ref {E29}), (\ref {E30}) and (\ref {E31}), (\ref {E32}).
It is seen that after the substitution
$ T_{s}^{a}(u) \to G_{-s}^{-a}(-u) $ and
$ G_{s}^{a}(u) \to T_{-s}^{-a}(-u) $ Eqs.
(\ref {E29}), (\ref {E30}) transform to
(\ref {E31}), (\ref {E32}).

In the case of zero boundary conditions
(\ref {E9}) for the function $ T_{s}^{a}(u)$,
the function $ G_{s}^{a}(u)$
satisfies also the same boundary conditions, but with
the maximum values $a$ less by a unity, i.e., $k$ should
be replaced by $(k-1)$. This feature expresses
Backlund transform which realizes the transform $A_{k} \to A_{k-1}$.
It was used in the paper \cite{Wieg} to get nested Bethe ansatz
equation. This transform for $k=3$ case can be
schematically presented in the following form:
\begin{equation}
\begin{array}{ccccccccccccc}
0&&1&&0&&&&&&&&\\
&&&&&&&&&&&&\\
0&&Q_1 (u+s)&& \bar Q_1 (u-s)&&0&&&&&&\\
&&&&&&&&&&&&\\
0&&Q_2 (u+s)&&G_{s}^{1}(u)&& \bar Q_2 (u-s)&&0&&&&\\
&&&&&&&&&&&&\\
0&&\phi (u+s)&&T_s^{1}(u)&&T_s^{2}(u)&& \bar \phi (u-s)&&0&&\\
\end{array}
\label{EE1}
\end{equation}
Here $Q_{\alpha}(u)$ are some boundary functions
expressed via Weierstrass functions $\sigma(\eta u))$,
$\bar Q_{\alpha}(u)= Q_{\alpha}(u-\alpha )$,
$ \bar \phi (u) = \phi (u-k)$.
It follows also from the boundary condition (\ref {E9})
that $Q_{k}(u)=\phi (u)$
(for details see Ref.[13]).

The functions at each level
satisfy Hirota equation. There is a relation between two
adjacent levels, defined by the equations of the linear problem.
The function $G_{s}^{1}(u)$ corresponds to the case
of $A_{1}$ algebra.  The equations of the linear problem
written for it allow to obtain integrals of motion,
which for $u$-dynamics \cite{Wieg} are:
\begin{eqnarray}
I_{1}(s) = \frac{Q_{2}(s-2)G_{s-u+1}^{1}(u) +
Q_{2}(s)G_{s-u-1}^{1}(u-2)}
{G_{s-u}^{1}(u-1)} \, ,
\label{EE2}
\\
\nonumber
\\
I_{2}(s) = \frac{\bar Q_{2}(s+2)G_{u-s+1}^{1}(u) +
\bar Q_{2}(s)G_{u-s-1}^{1}(u+2)}
{G_{u-s}^{1}(u+1)} \, .
\label{EE3}
\end{eqnarray}
For $A_{2}$ case Eq.(\ref {E29}) of the linear problem
for $a=1$ and $a=2$ may be written in the following way
\begin{eqnarray}
Q_{2}(u+s) T_{s+1}^{1}(u) - Q_{2}(u+s+2) T_{s}^{1}(u-1) =
c^{-1}G_{s+1}^{1}(u) \phi (u+s) \, ,
\label{EE4}
\\
\nonumber
\\
G_{s}^{1}(u) T_{s+1}^{2}(u) - G_{s+1}^{1}(u+2) T_{s}^{2}(u-1) =
c^{-1}\bar Q_{2}(u-s-1) T_{s}^{1}(u) \, .
\label{EE5}
\end{eqnarray}
Similarly Eq.(\ref {E30}) is written down
\begin{eqnarray}
G_{s}^{1}(u) T_{s+1}^{1}(u) - G_{s+1}^{1}(u-1) T_{s}^{1}(u) =
c Q_{2}(u+s+1) T_{s}^{2}(u-1) \, ,
\label{EE6}
\\
\nonumber
\\
\bar Q_{2}(u-s) T_{s+1}^{2}(u-1) - \bar Q_{2}(u-s-2) T_{s}^{2}(u) =
cG_{s+1}^{1}(u) \phi (u+s) \, .
\label{EE7}
\end{eqnarray}
Having expressed from (\ref {EE4}) the necessary
values $G_{s}^{1}(u)$ and substituting them in
(\ref {EE2}) we get the first integral of motion
$$
I_{1}(s) =\phi (s-2)\left [
\frac{Q_{2}(s-2)
\left[ T_{s-u+1}^{1}(u)Q_{2}(s) - T_{s-u}^{1}(u-1)Q_{2}(s+2)\right]
}
{\phi (s) \left[T_{s-u}^{1}(u-1)Q_{2}(s-2) -
T_{s-u-1}^{1}(u-2)Q_{2}(s)\right]} \, + \right.
$$
\\
\\
\begin{equation}
\left.\frac{Q_{2}(s)
\left[ T_{s-u-1}^{1}(u-2)Q_{2}(s-4) - T_{s-u-2}^{1}(u-3)Q_{2}(s-2)\right]}
{\phi (s-4)\left [T_{s-u}^{1}(u-1)Q_{2}(s-2) -
T_{s-u-1}^{1}(u-2)Q_{2}(s)\right ]} \right ] \: .
\label{EE8}
\end{equation}
Similarly having expressed from (\ref {EE5}) the
corresponding values $G_{s}^{1}(u)$ and
substituting them in (\ref {EE3}) we get the second
integral of motion for $u-$dynamics
$$
I_{2}(s) =\phi (-s-2)\left [
\frac{\bar Q_{2}(2-s)
\left[ T_{s+u+1}^{2}(u-1)\bar Q_{2}(-s) - T_{s+u}^{2}(u)\bar Q_{2}(-s-2)
\right]}
{\phi (s)\left[T_{s+u}^{2}(u)\bar Q_{2}(2-s) -
T_{s+u-1}^{2}(u+1)\bar Q_{2}(-s)\right]} \, - \right.
$$
\begin{equation}
\left.\frac{\bar Q_{2}(-s)
\left[ T_{s+u-1}^{2}(u+1)\bar Q_{2}(4-s) - T_{s+u-2}^{1}(u+2)\bar Q_{2}(s-2)
\right]}
{\phi (s-4)\left[T_{s+u}^{2}(u)\bar Q_{2}(2-s) -
T_{s+u-1}^{2}(u+1)\bar Q_{2}(-s)\right]}
\right ]
\label{EE9}
\end{equation}
In order to obtain
integrals of motion
for $s-$dynamics one should
use the other pair of equations - Eqs.(\ref {EE5}), (\ref {EE7}).
These integrals of motion are related with the
form (\ref {E9}) of boundary conditions. Here one of
them is associated with the boundary condition at the left
end of the segment $[0, k]$ and the other - at the right
end. The chosen example $k = 3$  illustrates the method
of obtaining integrals of motion from the integrals of
motion for $k$ less by a unity.

Now we consider quasiperiodic boundary conditions.
In order Eqs.(\ref {E29}), (\ref {E30}) and (\ref {E31}), (\ref {E32}),
are fulfilled, the function $ G_{s}^{a}(u)$ in the case of the
quasiperiodic boudary conditions (\ref {E19}) should
also satisfy the same boundary conditions.
Therefore there are no Backlund flows in this case
and the transition from $A_{k}$ algebra to $A_{k-1}$ algebra
does not take place. It follows from the condition,
that two functions $T_{s}^{a}(u)$ and $ G_{s}^{a}(u)$
satisfy one and the same Hirota equation with
equal boundary conditions, that they are proportional
to each other. The relations between the indices in
these functions are defined by the law of the transformation
of direct equations of the linear problem
(\ref {E29}), (\ref {E30}) into the inverse ones
(\ref {E31}), (\ref {E32}). The coefficient of proportionality
should retain Hirota equation invariant, i.e. it should
coincide with the gauge transformation (\ref {E6}).
Thus we may write down
\begin{equation}
G_{s}^{a}(u) = \chi _{1}(a+u+s) \chi _{2}(a-u+s)
\chi _{3}(a+u-s) \chi _{4}(a-u-s)
T_{-s}^{-a}(-u) \, .
\label{E34}
\end{equation}
We use Eq.(\ref {E34}) in (\ref {E29}), (\ref {E30}).
As the result we obtain
$$
\frac{\chi _{4}(a-u-s)}{\chi _{4}(a-u-s-2)} T_{-s}^{-a}(-u) T_{s+1}^{a}(u+1)
- \frac{\chi _{1}(a+u+s+2)}{\chi _{1}(a+u+s)}
T_{-s-1}^{-a}(-u-1) T_{s}^{a}(u) =
$$
\begin{equation}
\\
\\
c^{-1}\frac{\chi _{2}(a-u+s-2)}{\chi _{2}(a-u+s)} T_{-s}^{-a+1}(-u-1)
T_{s+1}^{a+1}(u) \, ,
\label{E35}
\end{equation}
\\
$$
\frac{\chi _{2}(a-u+s-2)}{\chi _{2}(a-u+s)} T_{-s}^{-a+1}(-u-1)
T_{s+1}^{a}(u-1) -
\frac{\chi _{3}(a+u-s-2)}{\chi _{3}(a+u-s)} T_{-s-1}^{-a+1}(-u) T_{s}^{a}(u) =
\\
$$
\begin{equation}
c\frac{\chi _{4}(a-u-s)}{\chi _{4}(a-u-s-2)}
T_{-s}^{-a}(-u) T_{s+1}^{a-1}(u) \, .
\label{E36}
\end{equation}

We use this form of equations to get integrals of motion. 
Here we take into account that quasiperiodic boundary 
conditions limit the type of the functions 
$\chi _{1}$ and $\chi _{3}$. They should be periodic with 
the period $2k$, while the functions $\chi _{2}$ and $\chi _{4}$ 
may be arbitrary. We single out the functions $\chi _{1}$ and 
$\chi _{3}$ from Eqs. (\ref {E35}), (\ref {E36}), 
having constructed the relations
$$
\frac{\chi _{1}(a+u+s+2)}{\chi _{1}(a+u+s)} =
\frac{\chi _{4}(a-u-s)
T_{-s}^{-a}(-u) T_{s+1}^{a}(u+1)}
{\chi _{4}(a-u-s-2)
T_{-s-1}^{-a}(-u-1) T_{s}^{a}(u)}-
$$
\\
\begin{equation}
- c^{-1}\frac{\chi _{2}(a-u+s-2)
T_{-s}^{-a+1}(-u-1)T_{s+1}^{a+1}(u)}
{\chi _{2}(a-u+s)
T_{-s-1}^{-a}(-u-1) T_{s}^{a}(u)} \, ,
\label{E37}
\end{equation}
\\
$$
\frac{\chi _{3}(a+u-s-2)}{\chi _{3}(a+u-s)} =
\frac{\chi _{2}(a-u+s-2)
T_{-s}^{-a+1}(-u-1) T_{s+1}^{a}(u-1)}
{\chi _{2}(a-u+s)
T_{-s-1}^{-a+1}(-u) T_{s}^{a}(u)} -
$$
\\
\begin{equation}
- c\frac{\chi _{4}(a-u-s)
T_{-s}^{-a}(-u)T_{s+1}^{a-1}(u)}
{\chi _{4}(a-u-s-2)
T_{-s-1}^{-a+1}(-u) T_{s}^{a}(u)} \, .
\label{E38}
\end{equation}

It follows from the periodicity of the functions $\chi _{1}$ and 
$\chi _{3}$ that the right parts of Eqs. (\ref {E37}), (\ref {E38}) 
are periodic with the period $2k$ with respect to 
three indices. After the replacement $u\to u-s$ in Eq.(\ref {E37}) 
and the replacement $u\to u+s$ in (\ref {E38}) the right part of 
these equations does not depend on the time $s$. 
Consequently, after this replacement of variables 
the right hand size of the equations is the 
integral of motion for $s-$dynamics. 
The presence of the arbitrary functions $\chi_{i}$ and the 
constant $c$ in the formulas for the integrals of 
motion shows the degrees of freedom up to which we can 
consider the restriction imposed by them.  

Similarly we may obtain 
the preserving values of $u-$dynamics, substituting 
$s\to s-u$ in Eq.(\ref {E37}) and $s\to s+u$ in Eq.(\ref {E38}). 
For simplicity we assume that $\chi _{2} = \chi _{4} = 1$ and 
we consider the case of $A_{1}-$algebra. For $a=0$ and $a=1$ 
from Eq.(\ref {E37}) we may get two invariant combinations of 
functions:
\begin{equation}
\frac{\chi _{1}(u+s+2)}{\chi _{1}(u+s)} =
\frac{
T_{-s}^{0}(-u) T_{s+1}^{0}(u+1)}
{
T_{-s-1}^{0}(-u-1) T_{s}^{0}(u)}-
c^{-1}\frac{
T_{-s}^{1}(-u-1)T_{s+1}^{1}(u)}
{
T_{-s-1}^{0}(-u-1) T_{s}^{0}(u)} \, ,
\label{E39}
\end{equation}
\\
\begin{equation}
\frac{\chi _{1}(u+s+3)}{\chi _{1}(u+s+1)} =
\frac{
T_{-s}^{1}(-u+2) T_{s+1}^{1}(u+1)}
{
T_{-s-1}^{1}(-u+1) T_{s}^{1}(u)}-
c^{-1}\frac{
T_{-s}^{0}(-u-1)T_{s+1}^{0}(u-2)(\lambda e^{\alpha})^{2}}
{
T_{-s-1}^{1}(-u+1) T_{s}^{1}(u)} \, .
\label{E40}
\end{equation} 
The combinations made of the functions $\chi_{3}$, which 
follow from (\ref {E38}), have the form
\begin{equation}
\frac{\chi _{3}(u-s-2)}{\chi _{3}(u-s)} =
\frac{
T_{-s}^{1}(-u-1) T_{s+1}^{0}(u-1)}
{
T_{-s-1}^{1}(-u) T_{s}^{0}(u)} -
c\frac{
T_{-s}^{0}(-u)T_{s+2}^{1}(u+2)}
{
T_{-s-1}^{1}(-u) T_{s}^{0}(u)} \, ,
\label{E41}
\end{equation}
\\
\begin{equation}
\frac{\chi _{3}(u-s-1)}{\chi _{3}(u-s+1)} =
\frac{
T_{-s}^{0}(-u-1) T_{s+1}^{1}(u-1)}
{
T_{-s-1}^{0}(-u) T_{s}^{0}(u)} -
c\frac{
T_{-s}^{1}(-u+2)T_{s+1}^{0}(u)}
{
T_{-s-1}^{0}(-u) T_{s}^{0}(u)} \, .
\label{E42}
\end{equation}
The Eqs.(49)-(54) are the main result of the paper. 

It is seen directly from these equations that after the 
substitution $u\to u-s$ $(u\to u+s)$ or $s\to s-u$ $(s\to s+u)$ 
in Eqs.(\ref {E39}),(\ref {E40}) ((\ref {E41}),\ref {E42})) 
we obtain in the right hand sizes of these equations 
the magnitudes which do not depend on $s$ (on $u$) after the change of 
these discrete variables. 

\section*{4. Discussion }

The preserving combinations (\ref {E39}) - (\ref {E42}) of 
the functions $T_{s}^{a}(u)$ may be used for 
studying discrete dynamics in sine-Gordon model. 
To find the solutions of the equation of motion in the 
discrete sine-Gordon model, Bethe-Ansatz method 
was used in paper \cite{Wieg1}. In the case of $A_{1}$-algebra 
for the function 
\begin{equation}
T_{s}^{0,1}(u) =
A_{s}^{0,1}e^{\mu(s) u}
\prod\limits_{j=1}^N \sigma
\left(\eta \left(u-z_{j}^{0,1}(s)\right)\right),
\label{E43}
\end{equation} 
where the index $a$ takes the values $0, 1$, the function 
$\mu (s)= \mu_{0}s + \mu_{1}$ ($\mu_{0}, \mu_{1}$ are the 
arbitrary constants), two functions $z_{j}^{0,1}(s)$ 
are the zero functions of Weierstrass 
function and are defined by the solutions of BA equations:
$$
\prod\limits_{j=1}^N
\left(
\frac{
\sigma(\eta (z_{i}^{1}(s+1)-z_{j}^{0}(s+1)-1))}
{\sigma(\eta (z_{i}^{1}(s+1)-z_{j}^{0}(s+2)-3))}
\right)
\left(
\frac{
\sigma(\eta (z_{i}^{1}(s)-z_{j}^{1}(s)+1))}
{\sigma(\eta (z_{i}^{1}(s+1)-z_{j}^{1}(s+1)+1))}
\right)
$$
\begin{equation}
\times \frac{
\sigma(\eta (z_{i}^{1}(s+1)-z_{j}^{0}(s+2)-2))}
{\sigma(\eta (z_{i}^{1}(s+1)-z_{j}^{0}(s)))} =
-\frac{A_{s}^{0}A_{s+1}^{1}}{A_{s+1}^{0}A_{s+1}^{1}}
\label{E44}
\end{equation}
$$
\prod\limits_{j=1}^N
\left(
\frac{
\sigma(\eta (z_{i}^{0}(s+1)-z_{j}^{1}(s+1)+2))}
{\sigma(\eta (z_{i}^{0}(s+1)-z_{j}^{1}(s+1)+1))}
\right)
\left(
\frac{
\sigma(\eta (z_{i}^{0}(s)-z_{j}^{0}(s)+1))}
{\sigma(\eta (z_{i}^{0}(s)-z_{j}^{0}(s)+1))}
\right)
$$
\begin{equation}
\times \frac{
\sigma(\eta (z_{i}^{0}(s+1)-z_{j}^{1}(s+2)-1))}
{\sigma(\eta (z_{i}^{0}(s+2)-z_{j}^{1}(s+2)))} =
-\frac{A_{s}^{0}A_{s+1}^{1}}{A_{s+1}^{0}A_{s+1}^{1}}
\label{E45}
\end{equation} 
In the rational limit, when $\sigma(\eta u)\to u$, 
the functions $A_{s}$ may be found from the 
system of equations
\cite{Wieg1}
\begin{equation}
(A_{s}^{1})^{2} - A_{s+1}^{1}A_{s-1}^{1} =
e^{\alpha }\lambda^{2}(A_{s}^{0})^{2}\equiv
a(A_{s}^{0})^{2}
\label{E46}
\end{equation}
\begin{equation}
(A_{s}^{0})^{2} - A_{s+1}^{0}A_{s-1}^{0} =
e^{-\alpha }\lambda^{2}(A_{s}^{1})^{2}\equiv
b(A_{s}^{1})^{2}
\label{E47}
\end{equation} 
The equations (\ref {E44}) - (\ref {E47}) describe the 
discrete dynamics of the zeroes $z_{j}^{0,1}(s)$ as the 
function of the time $s$.

In the presence of the restrictions following from 
(\ref {E39})-(\ref {E42}) to solve Eqs.(\ref {E44}), (\ref {E45}), 
one should find the solutions of Eqs.(\ref {E46}), (\ref {E47}). 
We pay attention to the invariance of Eqs.(\ref {E46}), (\ref {E47}) 
with respect to the transformation of 
$(A_{s}^{0}, A_{s}^{1}) \to (A_{s}^{0}, A_{s}^{1})e^{\delta s}$, 
where $\delta -$ is the constant. This invariance is already 
displayed in Eq.(\ref {E43}) with the aid of the functions 
$\exp (\mu (s)\,u )$ with $\mu = \mu_{0}\,s + \mu_{1}$.

It follows from Eqs.(\ref {E46}), (\ref {E47}) that the 
functions $A_{s}^{0,1}$ have the form
\begin{equation}
A_{s}^{0}=
Ae^{-\gamma s^{2}},
\label{E50}
\end{equation}
\begin{equation}
A_{s}^{1}=
Ae^{-\gamma s^{2}+\alpha/2},
\label{E51}
\end{equation} 
Here the coefficients $\gamma$ and $\lambda$ are 
connected by the relation $\lambda^{2}=1-\exp (-2\gamma)$, 
$A$ is the constant. Using the solutions obtained 
(\ref {E50}), (\ref {E51}) in the right hand size of Eqs.(\ref {E44}), 
(\ref {E45}) we get that the right hand size of these equations 
equals $-1$. 
In this case, we have to change certainly 
the functions in the left hand size 
of Eqs.(56),(57) by $\sigma(\eta u)\to u$. 

Substituting the solutions 
$T_{s}^{0,1}(u)$ of (\ref {E43}) to the gauge-invariant 
functions $Y_{s}^{a}(u)$ we see that taking into account 
(\ref {E50}) the dependence of these functions on $s$ is defined 
only by the dependence of the roots $z_{s}^{0,1}$ on the 
discrete time $s$ and does not depend on Gauss multipliers (56), (57). 
If we substitute the functions $T_{s}^{0,1}(u)$ from Eq.(\ref {E43}) 
to the limitations (\ref {E39})-(\ref {E42}) expressing 
the conservation laws in the considered system, 
this will allow to find the limitations on the dependencies 
of the roots $z_{s}^{0,1}$ on the discrete time $s$. 
In the case $N= 2$ this problem may be presented in the 
form of the finite number of mappings. The thermodynamic limit 
$N\to \infty$ requires numerical calculations.

In conclusion, we found the integrals of motion for $s-$ and $u-$dynamics 
for $A_{k}-$algebra with quasiperiodic boundary conditions. 
The case $k = 2$ corresponds to the discrete sine-Gordon model. 
It is shown that for $A_{2}-$ algebra the discrete 
equations of motion at zero boundary conditions 
convert in the continuous limit to the equations of 
motion of non-Abelian generalization of Liouville 
equation and at the quasiperiodic boundary conditions - 
to the equation of motion of the $SU(3)$ affine Toda equation and 
to the equation of motion of the Bullough-Dodd model.

\section*{Acknowledgements}

We want to thank A.I. Bobenko and W.K. Schief for useful 
discussion. This work was supported in part by the Russian Foundation 
for Basic Research under 
Grant No. 98-02-16237.

\end{document}